\documentclass{article}
\usepackage{spconf,amsmath,graphicx}

\usepackage{cite}
\usepackage{amsmath,amssymb,amsfonts}
\usepackage{algorithmic}
\usepackage{graphicx}
\usepackage{textcomp}
\usepackage[dvipsnames]{xcolor}
\usepackage{hyperref}
\usepackage{listings}
\usepackage{xcolor}
\usepackage{subcaption}
\usepackage{booktabs}

\usepackage{enumitem}
\setlist{nosep, leftmargin=14pt}

\usepackage{mwe} %

\definecolor{codegreen}{rgb}{0,0.6,0}
\definecolor{codegray}{rgb}{0.5,0.5,0.5}
\definecolor{codepurple}{rgb}{0.58,0,0.82}
\definecolor{backcolour}{rgb}{0.95,0.95,0.92}

\lstdefinestyle{mystyle}{
    backgroundcolor=\color{backcolour},   
    commentstyle=\color{codegreen},
    keywordstyle=\color{magenta},
    numberstyle=\tiny\color{codegray},
    stringstyle=\color{codepurple},
    basicstyle=\ttfamily\footnotesize,
    breakatwhitespace=false,         
    breaklines=true,                 
    captionpos=b,                    
    keepspaces=true,                 
    numbers=left,                    
    numbersep=5pt,                  
    showspaces=false,                
    showstringspaces=false,
    showtabs=false,                  
    tabsize=2
}

\newcommand{\std}[1]{\scriptsize{$\pm$#1}}

\title{Multi-target stain normalization for histology slides}
\name{Desislav Ivanov$^{1,3}$\sthanks{Work done while the author was at University of Turin. Now at Google.} \quad Carlo Alberto Barbano$^{1,2}$\sthanks{Correspondence: carlo.barbano@unito.it} \quad Marco Grangetto$^1$}
\address{$^1$University of Turin \quad $^2$LTCI, Télécom Paris, IP Paris \quad $^3$Google}
\begin{document}
\maketitle
\begin{abstract}
Traditional staining normalization approaches, e.g. Macenko, typically rely on the choice of a single representative reference image, which may not adequately account for the diverse staining patterns of datasets collected in practical scenarios. In this study, we introduce a novel approach that leverages multiple reference images to enhance robustness against stain variation. Our method is parameter-free and can be adopted in existing computational pathology pipelines with no significant changes.
We evaluate the effectiveness of our method through experiments using a deep-learning pipeline for automatic nuclei segmentation on colorectal images. Our results show that by leveraging multiple reference images, better results can be achieved when generalizing to external data, where the staining can widely differ from the training set. 
\end{abstract}

\section{Introduction}

\begin{figure}
    \centering
    \includegraphics[width=\columnwidth]{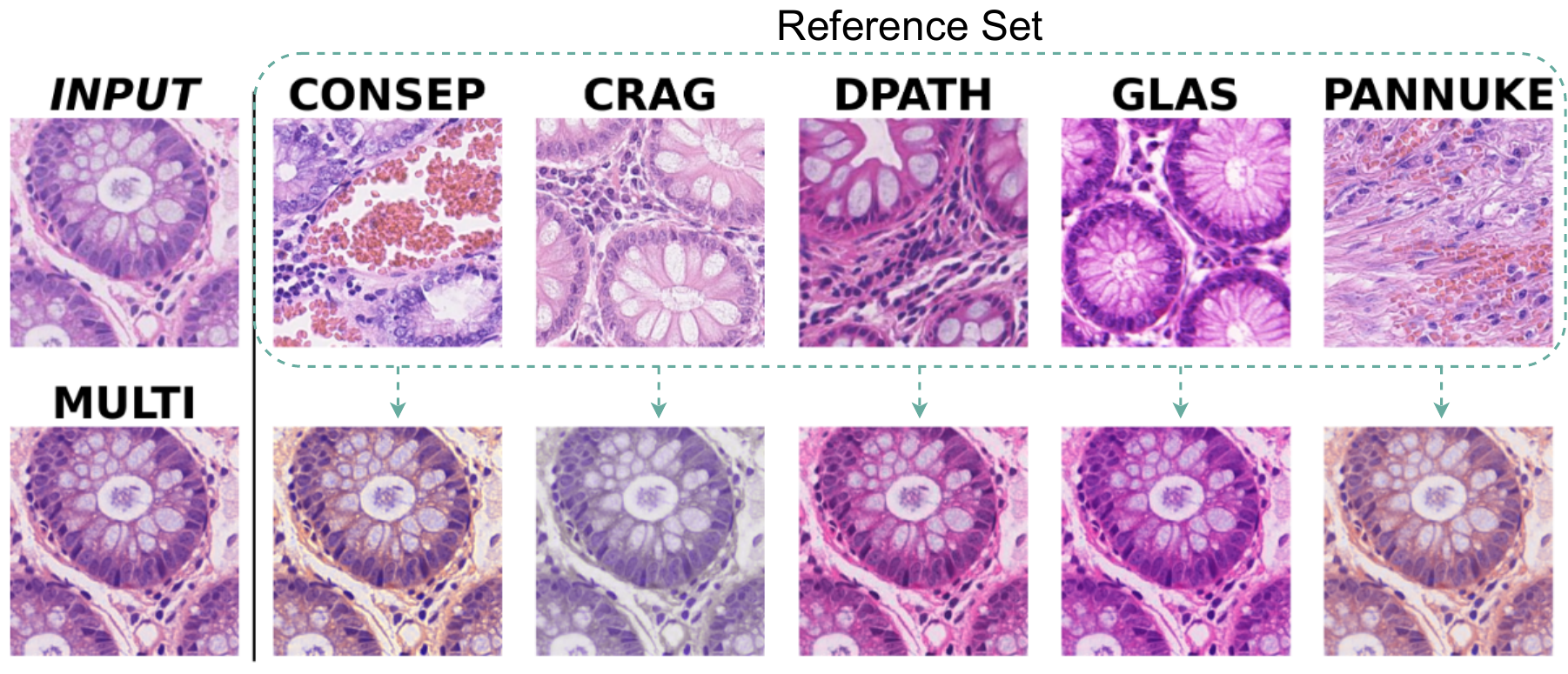}
    \caption{Multi-target normalization is more robust to stain variation than traditional approaches that consider only one target image. In the figure, a given input image is normalized with different targets (\emph{top row}) chosen from different sources, using the method in~\cite{macenko}. Multi-target (\emph{bottom left}) employs all of the reference images.}
    \label{fig:teaser}
\end{figure}

Stain normalization is a crucial step in computational pathology, as it helps to reduce the impact of variability in slide staining on downstream tasks for both pathologists and computer-aided-diagnosis systems~\cite{michielli}. Stain normalization is particularly important because histopathological images are often stained with different dyes or staining techniques, such as Hematoxylin and Eosin (H\&E) which are the most commonly used stains~\cite{chan}, leading to variations in intensity and color across different slides. These variations can make it challenging to identify features of interest, such as tumor cells or tissue structures, and can limit the accuracy of automated analysis algorithms.
Several methods have been proposed in the literature for stain normalization~\cite{macenko, reinhard, roy, vahadane}, but most of them rely on the choice of a single representative reference image, which may not adequately account for the diverse staining patterns present in datasets collected in practical scenarios. This can lead to suboptimal performance and reduced robustness against stain variation.

In this study, we introduce a novel approach to stain normalization that leverages multiple reference images to enhance robustness against stain variation. Our method is parameter-free, meaning it does not introduce any additional complexity into existing computational pathology pipelines. By computing a reference stain matrix for different reference image, we more accurately capture the underlying patterns in the data and achieve better robustness against stain variation.
We validate our approach empirically, using a deep-learning pipeline for automatic nuclei segmentation on colorectal images. Our results show that by leveraging multiple reference images, better results can be achieved when generalizing to external data, where the staining can widely differ from the training set. 
We analyze different possible formulations for including multiple reference images and find that averaging the reference stain matrix of each image is best in terms of reliability and robustness of the normalization.
The proposed method has the ability to improve the accuracy and robustness of downstream tasks, e.g. nuclei segmentation, and can easily be applied to other computational pathology tasks.
\section{Materials and Methods}

\subsection{Experimental data}
We employ the Lizard dataset\cite{lizard}, which consists of H\&E stained histological images with their respective nuclei segmentation and classification. The dataset is a collection of 6 publicly available datasets: GlaS, CRAG, CoNSeP, DigestPath, PanNuke, and TCGA~\cite{lizard}. Lizard provides 291 high-resolution images, that include multiple organs. In this work, we focus our analysis on colorectal samples. Each nucleus is classified into one of 6 different classes: Epithelial, Lymphocyte, Plasma, Neutrophil, Eosinophil, and Connective. 

\subsection{Proposed method}
\label{sec:method}

\subsubsection{Background}
The detected intensities of light transmitted through a
specimen with the concentration of stain ($c$) is described by Lambert-Beer’s law~\cite{lambertbeer}:

\begin{equation}
    I = I_{0} \exp(-v c)
    \label{eq:lambertbeer-law}
\end{equation}

\noindent where $I_0$ is the intensity of light entering the specimen, $I$ is the intensity of the detected light, and $v = \varepsilon\ell$ where $\varepsilon$ is the stain absorption coefficient and $\ell$ is the optical path length.
As we can see from Eq.~\eqref{eq:lambertbeer-law}, the intensity $I$ picked up by the camera depends on the stain concentration non-linearly. A linear dependency can be obtained, as standard practice~\cite{macenko}, by converting the values to optical density (OD):

\begin{equation}
    I_{OD} = -\log_{10}(I/I_0) = VS
\end{equation}

\noindent where $V$ and $S$ are the stain matrices and the concentration of each of the stains, respectively. From here, once $V$ is determined by some method, it is possible to retrieve the stain concentration $S$ with a simple de-convolution scheme as in~\cite{ruifrok}:

\begin{equation}
    S = V^{-1} I_{OD}
\end{equation}

\noindent In~\cite{ruifrok}, a method to determine $V$ is proposed by acquiring samples stained with only a single stain (H or E). 
In~\cite{macenko} a more robust and automatic method for determining the stain matrix is proposed, based on the SVD of the OD matrix. First, OD values below a certain threshold are discarded (they correspond to empty pixels); in~\cite{macenko} a threshold $\beta = 0.15$ is used. Then, SVD is computed on the remaining OD values, and the plane formed by the two vectors corresponding to the two largest singular values is considered. All of the OD values are projected onto this plane and normalized to unit length. For each projected vector, the angle $\phi$ wrt to the first SVD direction is computed, and two robust extremes are found as the $\alpha^{th}$ and $(100 - \alpha)^{th}$ percentiles, in \cite{macenko} good results are obtained with $\alpha=1$. These two extreme values are then converted back to OD space and correspond to the estimated stain vectors.

In order to perform stain normalization with respect to a reference image, the reference stain matrix $V_{ref}$ is computed, and it is then used to reconstruct the normalized image from the computed concentration matrix $S$:

\begin{equation}
    I_{OD}^{norm} = V_{ref} S 
\end{equation}

\subsubsection{Multi-target normalization}
In~\cite{macenko} only a single reference image is used to estimate the stain matrix $V_{ref}$. We propose to make this estimation more robust, by computing it across a set of reference images. Given a set $\mathcal{T}$ of reference images, there are different possible ways to do this:

\begin{enumerate}
    \item \textbf{Stochastic:} pick a random reference image $t \in \mathcal{T}$ and use its reference matrix $V_{ref}^t$ for normalization every time a new image has to be normalized. In Fig.~\ref{fig:teaser} this would correspond to considering a random image in the bottom row;

    \item \textbf{Concat:} concat all reference images into a large single image $T$ and compute its reference matrix $V_{ref}^{T}$; the rest of the algorithm is unchanged;

    \item \textbf{Avg-pre:} compute the main SVD directions for each reference image and average them. To compute the robust extremes, project all the OD values of the target images using the average SVD direction;
    
    \item \textbf{Avg-post:} Compute the reference matrix $V_{ref}^t$ for each reference image $t \in \mathcal{T}$ and use the average $\mathbb{E}[V_{ref}^t]$ as final reference matrix.  %
\end{enumerate}

\section{Results}

To evaluate our proposed normalization method, we employ a deep-learning-based pipeline for automatic segmentation.

\subsection{Experimental setup}
\label{sec:setup}
We use a UNet++ model with a ResNet18 backbone pretrained on ImageNet. To train the models, we use the Adam optimizer for 150 epochs with a learning rate of 0.001, a batch size of 64, betas of 0.9 and 0.999 respectively, and no weight decay. Our implementation is based on PyTorch~\cite{pytorch}, and the experiments were run on the Leonardo\footnote{We acknowledge the CINECA award under the ISCRA initiative, for the availability of high performance computing resources and support. This work was granted access to HPC resources under the allocation IsB28\_HyGenAI.} cluster on NVIDIA A100 GPUs. The normalization algorithms implementation is based on the torchstain library~\cite{torchstain}.

\begin{figure*}
  \centering
  \begin{subfigure}[b]{0.32\linewidth}
    \includegraphics[width=\linewidth]{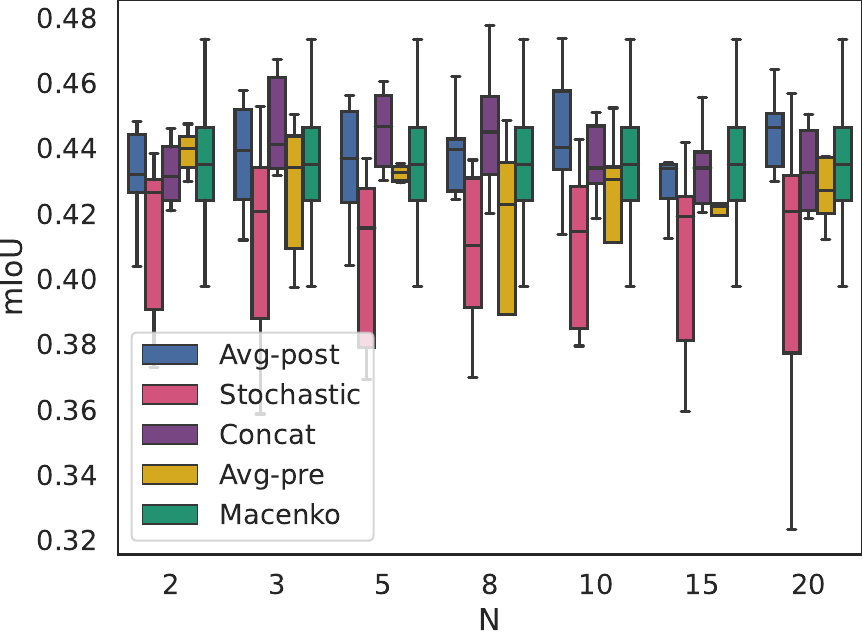}
    \caption{~}
    \label{fig:results-same}
  \end{subfigure}
  \begin{subfigure}[b]{0.32\linewidth}
    \includegraphics[width=\linewidth]{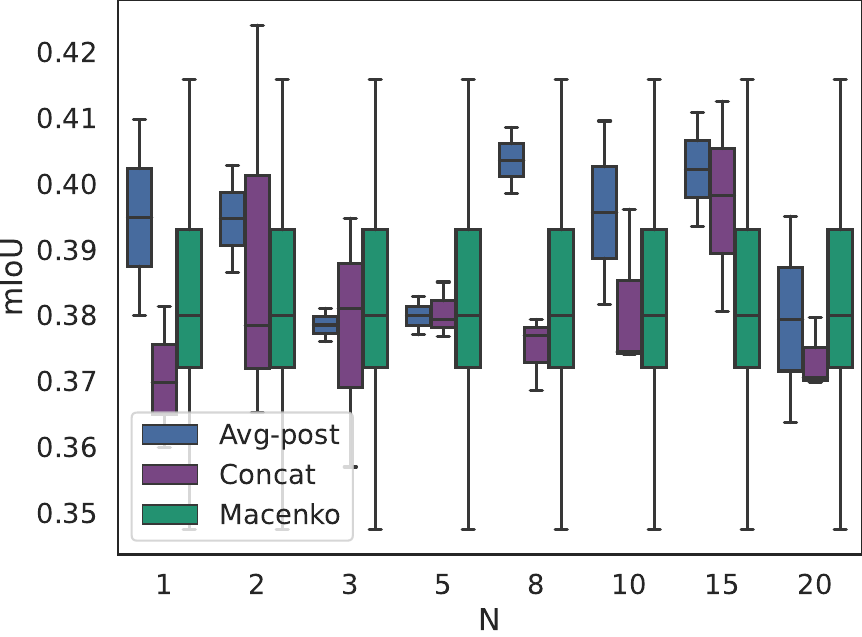}
    \caption{~}
    \label{fig:transfer-nonorm}
  \end{subfigure}
  \begin{subfigure}[b]{0.32\linewidth}
    \includegraphics[width=\linewidth]{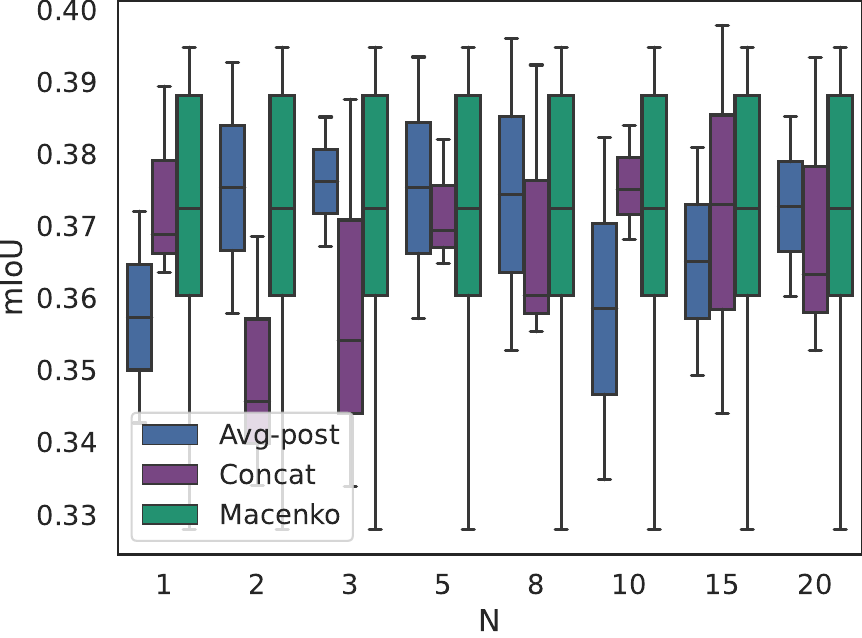}
    \caption{~}
    \label{fig:transfer-norm}
  \end{subfigure}
  \caption{Comparison of the mIoU (y-axis) between different normalization methods based on the size of the subset of reference images (x-axis). Training and testing on the whole Lizard dataset (a) Generalization to external data w/o normalization during training (b) Generalization to external data w/ Macenko normalization during training (c).
  }
  \label{fig:results-main}
\end{figure*}

\subsection{Evaluation on same dataset}
\label{sec:results-same}

\begin{figure}
    \centering
    \includegraphics[width=\columnwidth]{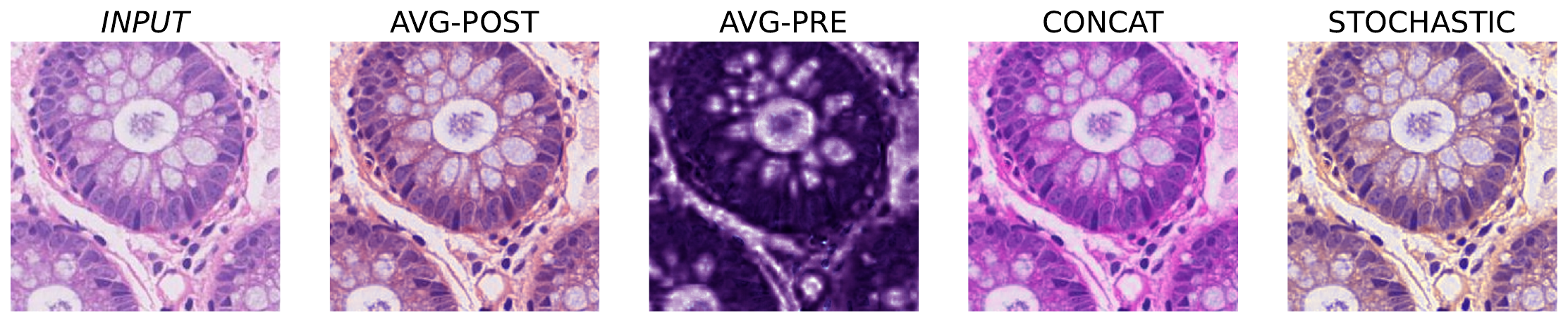}
    \caption{Visual comparison of the different multi-target approaches.}
    \label{fig:norm-comparison}
\end{figure}

As a first benchmark, we employ the full Lizard dataset for assessing the impact of the different normalization techniques. To this end, we train a deep segmentation model with the setup described in Sec.~\ref{sec:setup}. We adopt a random train-test split of the data, with 80\% of the data for training and 20\% for testing. We investigate the results varying the size $N$ of the reference set containing the target images for normalization. In our experiments, we vary $N$ between 2 and 20. To build the reference set $\mathcal{T}$, we randomly select $N$ images from the training set. For each of the proposed methods we run three independent trials and we evaluate the mean intersection-over-union (mIoU) performance on the test set. 
The results are presented in Fig.~\ref{fig:results-same}. As a baseline, we report the results of the standard Macenko normalization~\cite{macenko}. The worst results are obtained by the \emph{stochastic} method. The mIoU is always lower than the baseline, denoting that high variability in staining references during training is detrimental to model convergence. This is also shown by the large standard deviation of the results. In fact, as shown in Fig.~\ref{fig:teaser}, the normalization output can widely differ depending on the chosen target image. The Macenko baseline achieves on average higher results, meaning that in terms of convergence it is better to pick a single image and stick with it for the whole training. Slightly better results are achieved by the \emph{avg-pre} method, however, the average mIoU is still lower than the baseline most of the times. From these results, we conclude that averaging the main SVD components of different images is a suboptimal choice as it may introduce artifacts in the computation of the robust extremes, due to the widely different OD values of the images. The best results are achieved by the \emph{concat} and \emph{avg-post} methods, peaking at $N=8$ and $N=10$ respectively. Also, both methods achieve average results higher than the baseline in most cases. The overall results are summarized in Tab.~\ref{tab:results-same}.
\begin{table}
    \centering
    \resizebox{\linewidth}{!}{
    \begin{tabular}{c c c c c c}
    \toprule
         &  \textbf{Macenko} & \textbf{Stochastic} & \textbf{Avg-Pre} & \textbf{Concat} & \textbf{Avg-Post} \\
    \midrule
    \textbf{mIoU} &  0.43\std{0.02} & 0.37\std{0.10} & 0.42\std{0.01} & 0.43\std{0.03} & \textbf{0.44 \std{0.01}}\\
    \bottomrule
    \end{tabular}
    }
    \caption{Results in terms of mIoU on Lizard.}
    \label{tab:results-same}
\end{table}
For further analysis, in Fig.~\ref{fig:norm-comparison} we visually show the results of the different multi-target normalization methods. The reference set is the same as of Fig.~\ref{fig:teaser}. First of all, we notice that \emph{avg-pre} actually introduces artifacts in the reconstruction, which confirms the results previously presented. For \emph{stochastic}, the results in terms of normalization are definitely better, however, the considerations made about Fig.~\ref{fig:teaser} must be taken into account. \emph{Concat} and \emph{avg-post}, which obtain the best results, exhibit noticeable differences in the reconstructed staining. In fact, one could argue that by simply concatenating the input image, the most predominant staining concentration in the reference set will mainly drive the normalization process. This can be observed by comparing the normalization output with the reference images in Fig.~\ref{fig:teaser}, as the GLAS and DPATH samples exhibit the strongest colorization. On the other hand, \emph{avg-post} strikes a favorable balance between all the reference stain matrices. For this reason, we expect \emph{avg-post} to be more reliable in practical applications.

\subsection{Generalization to external data}

To evaluate our claims, we now assess how the proposed methods can help generalization to novel data. To simulate this occurrence, we train a segmentation model on the training set, excluding one dataset, which is kept only for testing. We then measure the mIoU on the excluded dataset, picking the reference images from the training set. We study two settings, one in which the model is trained without any normalization, and one in which it is trained with Macenko normalization. In both cases, the reference images are picked at random and do not necessarily include the one(s) used during training. This experimental setting is relevant for real-world scenarios, in which a pre-trained model, e.g. publicly available checkpoint, can be leveraged but knowledge about the normalization method or the reference set is not provided. The results are shown in Fig.~\ref{fig:transfer-nonorm} and~\ref{fig:transfer-norm}. In both cases, we achieve improvements with respect to the baseline (Macenko with random reference image from the training set). Furthermore, \emph{avg-post} achieves more consistent results than \emph{full-concat}, as hypothesized in Sec.~\ref{sec:results-same}.

\section{Software Implementation}

We make our implementation available in the torchstain\footnote{\url{https://github.com/EIDOSLAB/torchstain}} normalization library. Our method is contained in the \emph{MultiMacenkoNormalizer} class. A code snippet is presented below.

\begin{lstlisting}[language=Python]
from torchstain.normalizers import *
... 

ref_images = sample_random_images(dataset, N)

normalizer = MultiMacenkoNormalizer('avg-post')
normalizer.fit(ref_images)

img = next(iter(dataset))
norm, _, _ = normalizer.normalize(img)
\end{lstlisting}

\section{Related Works}
Several methods have been proposed for stain normalization in computational pathology. Among the most widely used, we can find the work by Macenko et al.~\cite{macenko}, which is presented in Sec.~\ref{sec:method}. Other de-facto standard methods include Reinhard normalization~\cite{reinhard}, which consists of a color transfer technique based on the $l\alpha\beta$ space, and the modified version~\cite{roy}, which has been specifically developed to solve some of its limitations and was applied to histopathology images. Another very common stain normalization method is introduced in~\cite{vahadane}. This work presented a structure-preserving color normalization method that separates the stains into sparse density maps. More recently, approaches based on deep learning have been proposed. One of the most popular approaches is StainNet~\cite{stainnet}, which is based on generative adversarial networks (GAN), in order to transform an image from one domain to another. Of course, the latter kind of approaches introduce a lot of complexity and their applicability may be often limited to deep learning practitioners.

\section{Conclusions}

In this work, we introduced a novel stain normalization technique for histology slides, which is capable of leveraging multiple reference images for the normalization. Our method, differently from other existing works, is parameter-free and thus does not introduce any additional complexity in computational pathology pipelines. Our experiments show that by leveraging multiple reference images, better results can be achieved when generalizing to external data, where the staining can widely differ from the training set. Our method is based on the widely used stain normalization algorithm proposed by Macenko~\cite{macenko}. We analyze different formulations for including multiple reference images, and we empirically find that averaging the reference stain matrix of each image is the best choice in terms of reliability and robustness of the normalization. \\ %

\noindent \textbf{Compliance with ethical standards} This research was conducted retrospectively using
human subject data made available in open access by~\cite{lizard}. Ethical approval was not required as confirmed by the license attached with the data.

\end{document}